\documentclass[journal = jpclcd, manuscript=letter]{achemso}
\setkeys{acs}{articletitle = false}
\usepackage{amsmath}
\usepackage{amsfonts}
\usepackage{amssymb}
\usepackage{subcaption}
\usepackage{graphicx}

\newcommand{\ave}[1]{\left\langle #1 \right\rangle}

\author{Michael McEldrew}
\affiliation[MITCHE]
{Department of Chemical Engineering, Massachusetts Institute of Technology, Cambridge, MA, USA}

\author{Zachary A. H. Goodwin}
\affiliation[IMPPhys]
{Department of Physics, CDT Theory and Simulation of Materials, Imperial College of London, South Kensington Campus, London SW7 2AZ, UK}
\alsoaffiliation[IMPChem]
{Department of Chemistry and Thomas Young Centre for Theory and Simulation of Materials, Imperial College of London, South Kensington Campus, London SW7 2AZ, UK}

\author{Alexei A. Kornyshev}
\email{a.kornyshev@imperial.ac.uk}
\affiliation[IMPChem]
{Department of Chemistry, Imperial College of London, South Kensington Campus, London SW7 2AZ, UK}
\alsoaffiliation[TYC]
{Thomas Young Centre for Theory and Simulation of Materials, Imperial College of London, South Kensington Campus, London SW7 2AZ, UK}

\author{Martin Z. Bazant}
\email{bazant@mit.edu}
\affiliation[MITCHE]
{Department of Chemical Engineering, Massachusetts Institute of Technology, Cambridge, MA, USA}
\alsoaffiliation[MITMATH]
{Department of Mathematics, Massachusetts Institute of Technology, Cambridge, MA, USA}
\title{Theory of the Double Layer in Water-in-Salt Electrolytes}

\begin{document}

\section{Abstract}
One challenge in developing the next generation of lithium-ion batteries is the replacement of organic electrolytes, which are flammable and most often contain toxic and thermally unstable lithium salts, with safer, environmentally friendly alternatives. Recently developed Water-in-Salt Electrolytes (WiSEs) were found to be a promising alternative, having also enhanced electrochemical stability. In this work, we develop a simple modified Poisson-Fermi theory, which demonstrates the fine interplay between electrosorption, solvation, and ion correlations. The phenomenological parameters are extracted from molecular simulations, also performed here. The theory reproduces the electrical double layer structure of WiSEs with remarkable accuracy.
\newpage

\section{Main Text}
Recently, Water-in-Salt Electrolytes (WiSEs) have emerged as promising replacements for organic electrolytes in alkali-ion batteries \cite{Suo2013,Suo2015,Smith2015,Wang2016,Wang2018,Wang2018a,Sun2017,Sodeyama2014,Yamada2016,kuhnel2017,leonard2018} with dramatically improved safety and stability and potentially lower cost. These ultra-concentrated aqueous electrolytes display enhanced electrochemical stability windows (ESWs), potentially enabling lithium based WiSEs to operate up to 4 V \cite{yang2017}. The reductive stability of WiSEs has been largely attributed to their ability to form a passivating solid-electrolyte interface (SEI) at the anode \cite{Suo2015}, while oxidative stability has been thought to originate solely from the reduction of the thermodynamic activity of water. However, recent molecular dynamics (MD) simulations \cite{Vatamanu2017} have shown that water is practically depleted from positively biased electrode surfaces and presumably unable to react, which could contribute to the observed enhancement of the oxidative stability \cite{Vatamanu2017}. Interestingly, the response of water to positive and negative electrode biases is extremely asymmetric, and it is opposite to the trends observed for water in ionic liquids \cite{Feng2014}.
Although various hypotheses for the physical origin of this asymmetry have been put forward, here we present the first mathematical theory that is able to rationalize the behavior of WiSEs at electrified interfaces. 
\par
There is a vast literature on models of the electric double layer (EDL) of electrolytes~\cite{Bazant2009a,Fedorov2014}, much of which addresses the extremes of the concentration spectrum, especially the dilute limit. Here, the inspiration for modeling WiSEs comes predominantly from ionic liquids (ILs), solvent-free electrolytes in their pure form, which have been studied extensively \cite{Kornyshev2007,Fedorov2014,Bazant2011,Han2014,Maggs2016,Goodwin2017a,gavish2016,Blossey2017,Goodwin2017,Chen2017}. For ILs, finite size effects \cite{Kornyshev2007} (as in concentrated electrolytes~\cite{Bazant2009a,Kilic2007a,Bikerman1942}) and correlations \cite{Bazant2011, Goodwin2017a} play a critical role in the structure and thermodynamic response of the EDL. ILs, however, often contain water or other solvent additives, which are intentionally added to enhance transport properties \cite{seddon2000} or absorbed inadvertently from the environment \cite{wasserscheid2008}. Such systems have been modeled at the continuum level quite successfully with explicit account of the solvent molecules in modified Poisson-Fermi theories \cite{abrashkin2007,Budkov2015a,budkov2016,BBKGK,Chen2018}. 
\par
\begin{figure*}[!hbt]
    \centering
    \includegraphics[clip, width=1\textwidth]{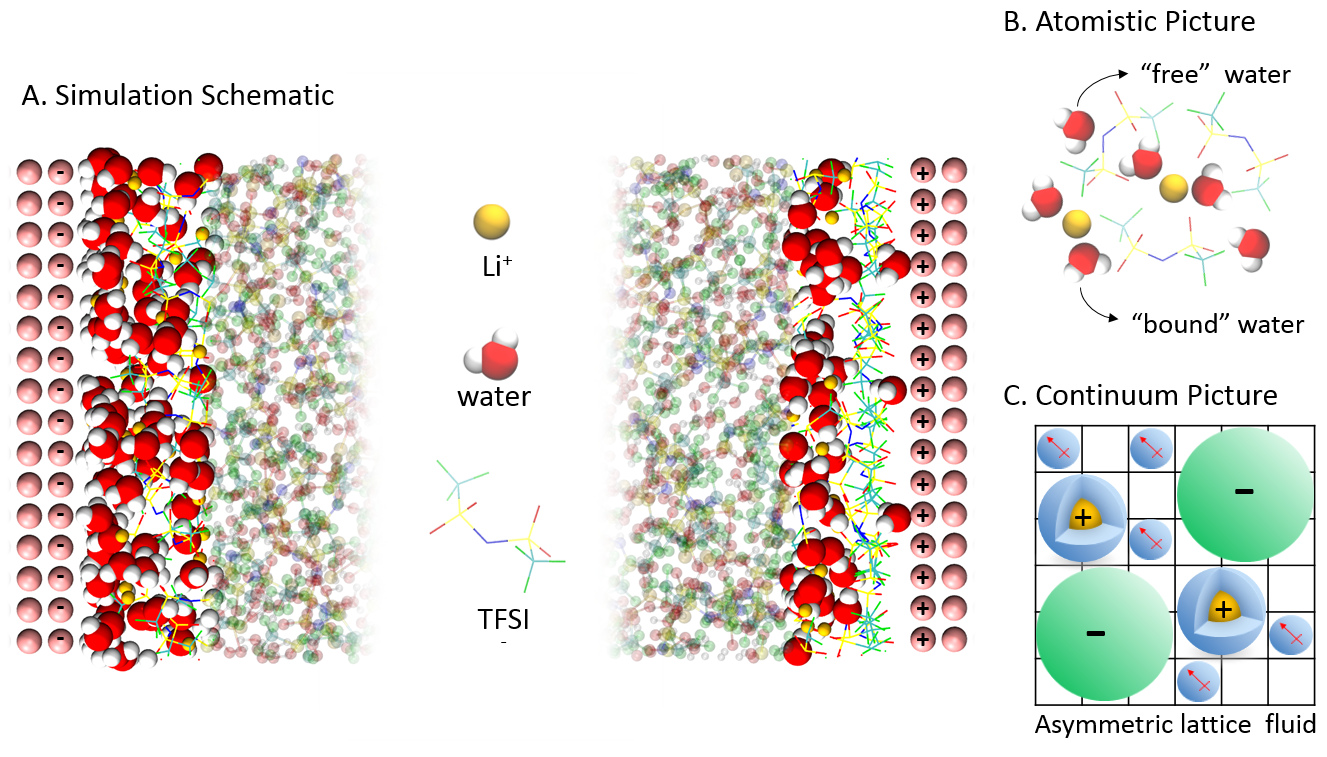}
    \caption{A) A schematic of molecular dynamics simulations performed for 21m lithium bis(trifluoromethane sulfonyl)imide (LiTFSI), which are used to parameterize and validate our novel continuum-level theory of the electrical double layer in WiSEs. Simulation details are given in the Supporting Information (SI). B) An atomistic representation of the fluid structure in 21m LiTFSI, detailing the two chemical states of water. C) An outline of the continuum 3-component (anion, solvated cation, and free water) lattice fluid description of the example WiSE. }
    \label{fig:md_box}
\end{figure*}

Here, we take a similar approach for WiSEs, but in our model the chemical state of water in the fluid is further developed conceptually. One remarkable point of the our theory is that all but one of the parameters were calculated using data from the performed MD simulations of an example WiSE, 21m lithium bis(trifluoromethane sulfonyl)imide (LiTFSI), with practically quantitative agreement between the simulations and theory (see fig. \ref{fig:md_box}A for schematic of MD simulations). 
\par
\par
WiSEs contain small Li$^+$ cations (or Na$^+$ and K$^+$in sodium and potassium-ion battery applications \cite{kuhnel2017,leonard2018}) and bulkier, fluorinated anions. For this reason, there is a large asymmetry in the water-ion interactions: the cation is strongly solvated, and the anion is not. The strong correlations between cations and water permits the partitioning of water into \emph{bound} and \emph{free} states, as outlined in fig. \ref{fig:md_box}B. Bound water molecules interact strongly with the cations and are unable to polarize freely in an external electric field. Thus, they contribute only marginally to the permittivity of the fluid and only appear implicitly in the model by adding to the molecular volume of cations. On the other hand, owing to their small concentration, free water molecules behave as fluctuating Langevin dipoles.

We model the WiSE as a dielectric medium, with background permittivity $\varepsilon_0\varepsilon_{\text{s}}$, in which fluctuating Langevin dipoles and solvated ions are embedded. The ions are correlated in the Bazant-Storey-Kornyshev (BSK) approximation \cite{Bazant2011}, and no ion pairing is considered \cite{Goodwin2017a,Chen2017}. In the Supporting Information (SI), the electrostatic energy density of the system, $w_e$, is derived. Note that our approach is qualitatively different from the approaches considering dielectric decrement from the dilute solution limit, where ions are taken to decrease the permittivity of bulk water due to local dielectric saturation \cite{Ben-Yaakov2009,Ben-Yaakov2011a,Ben-Yaakov2011b,lopez2011,hatlo2012,nakayama2015}. Our approach extends from the limit of no free water molecules, with  each additional free water molecule contributing to the permittivity in a Langevin-type manner \cite{Gongadze2011,Gongadze2011a,Gongadze2012,Gongadze2013,Gongadze2015,Budkov2015a,budkov2016,BBKGK}.

In experiments, the EDLs of WiSEs are generally in contact with thermal and chemical reservoirs. Thus, we assume the grand canonical free energy functional to be:
\begin{align}
\Omega =& \int_V d\mathbf{x}\big[w_e+f-\sum_i c_i \mu_i\big] 
\label{eq:BSK}
\end{align}
where $f(\{c_i\})$ is the free energy density, which is a function of the concentration of all species, $\{c_i\}$; and $\mu_i$ is the chemical potential of species $i = \pm,w$ fixed by the chemical reservoir of ions and water, respectively.

\par
We apply the Legendre transform to introduce the thermodynamic pressure, $p$ \cite{budkov2016,BBKGK}, consider a flat planar electrode with no variations parallel to the surface, plug in the explicit form of $w_e$ defined in the SI, and thus simplify Eq. \eqref{eq:BSK}, giving:
\begin{align}
\Omega =& - A \int dx \bigg[\frac{\varepsilon_0\varepsilon_s}{2}\left(|\phi'|^2+l_c^2|\phi''|^2\right) +p(\mu_+-e\phi,\mu_-+e\phi,\mu_w+\Psi)\bigg] 
\label{eq:BSK_G_new}
\end{align}
where $\phi$, $-\phi'$, $e$, $l_c$ and $A$ are the electrostatic potential, electric field, elementary charge, correlation length and cross sectional area, respectively. $\Psi$ stands for the energy of fluctuating dipoles in an electric field as $\Psi = k_BT \ln \left(\sinh(\beta p_w |\phi'|)/\beta p_w |\phi'| \right)$, with $\beta=1/k_BT$ and $p_w$ respectively denoting inverse thermal energy and dipole moment of free water.
\par
We model the WiSE as a 3 component asymmetric lattice fluid (see fig. \ref{fig:md_box}C), the pressure of which can be approximated with the explicit function:
\begin{align}
    p &= \frac{k_B T}{v_w}\ln \bigg[\Big(\left(1+\xi_+e^{\beta(\mu_+-e \phi(r))} \right)^{\frac{\xi_-}{\xi_+}}+ \xi_-e^{\beta(\mu_-+e\phi(r))}  \Big)^{\frac{1}{\xi_-}}+e^{\beta(\mu_w+\Psi(r))} \bigg]
    \label{eq:pressure}
\end{align}
where $\xi_\pm=\frac{v_\pm}{v_w}$. See the SI for a full derivation of Eq.\ref{eq:pressure}, which has been cast into an explicit form here for the first time to our knowledge. Our model assumes that free water molecules occupy a single lattice site, solvated cations occupy $\xi_+$ lattice sites, and anions occupy $\xi_-$ lattice sites.

The volume ratios, $\xi_-$ and $\xi_+$, are determined subject to the extent of bound water. $\text{Li}^+$ is strongly hydrates by water molecules, even at the extreme ionic concentrations in WiSEs, but the anions are significantly less so. Hence, although the `naked' $\text{Li}^+$ is significantly smaller than the anions, the hydrated cations are more comparable in size.
\par
\begin{figure}[!hbt]
    \centering
    \includegraphics[clip, trim=12cm 3cm 7cm 4cm , width=0.5\textwidth]{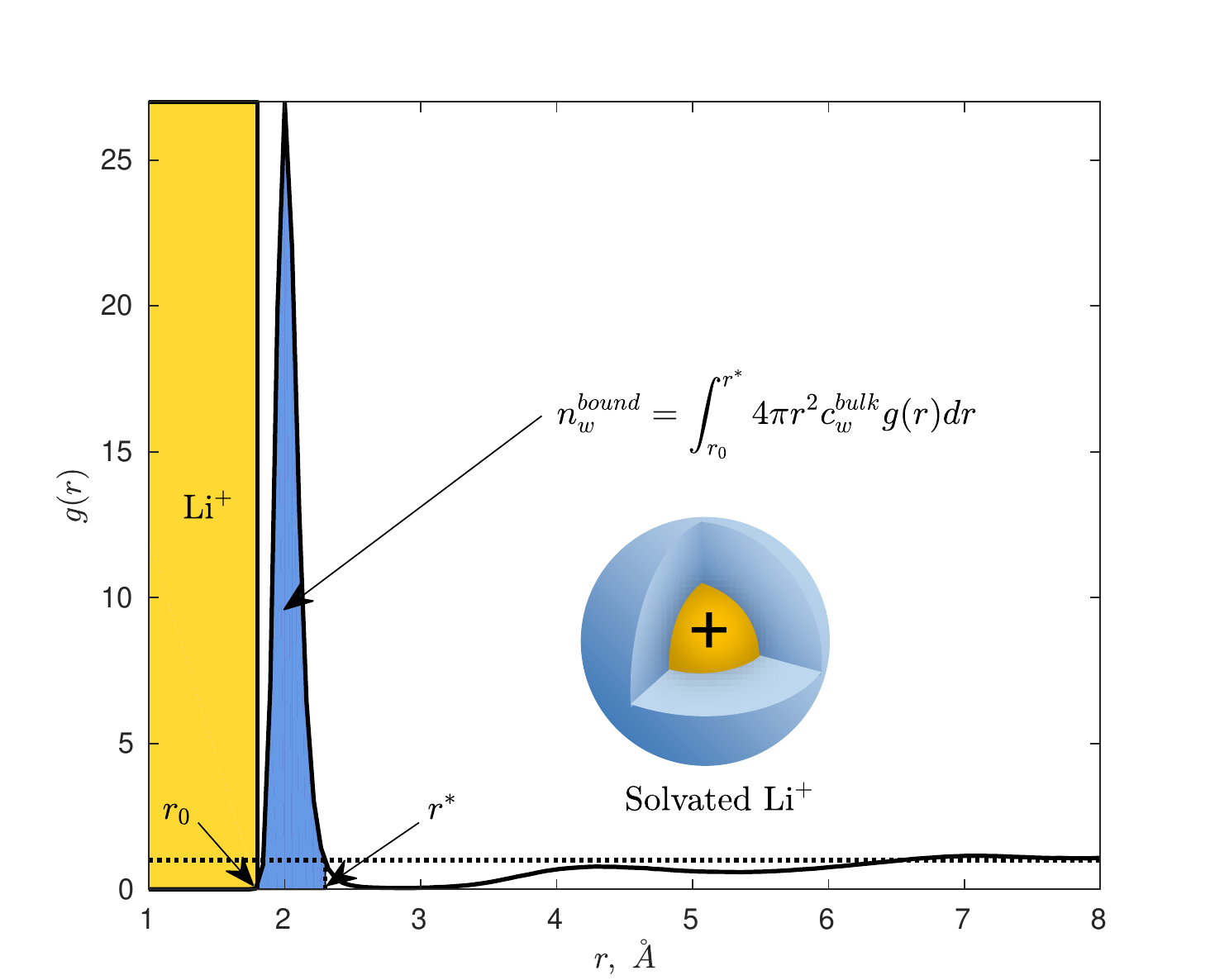}
    \caption{The Li-O pair correlation function. The shaded region is considered to be the first solvation layer of water around Li. These water molecules are considered to be bound to Li, forming a `hydrated Li$^+$ aquacomplex'.}
    \label{fig:rdf}
\end{figure}
\par
The extent of $\text{Li}^+$ hydration can be observed from the $\text{Li}^+-\text{O}$ pair correlation function in fig. \ref{fig:rdf}, which was computed from a MD simulation of 21m LiTFSI. We assume all water molecules within the first shell of $\text{Li}^+$ to be \emph{bound} to $\text{Li}^+$ and thus contribute to its molecular volume. The volume ratio of the cation to a water molecule is given by the volume of the $\text{Li}^+$ aquacomplex: $\xi_+ = \frac{4\pi(r^*)^3}{3v_w}$; where $r^*$ is the radius of the solvated $\text{Li}^+$ ion, as shown in fig \ref{fig:rdf}. For 21m LiTFSI, we find that for every $\text{Li}^+$ ion there are, on average, $n_w^{\text{bound}} = 2.5$  water molecules immediately bound to it. 
\par
To a first approximation we assume that the bulkier, more hydrophobic anions have no bound water molecules; thus, the anion to water volume ratio, $\xi_-$, is equal to the `naked' anion volume over the volume of a water molecule. This approximation was confirmed upon calculating the pair correlation function between water and the anion, as displayed in the SI, which showed a negligible hydration shell when compared to that of $\text{Li}^+$. The molecular volume of the naked anion was determined by computing the volume occupied by the overlapping Van der Waals radii of its atomic species. 
\par

The concentrations of species within the EDL may be determined directly from the pressure function using the thermodynamic relation, $c_i=\frac{\partial p}{\partial \mu_i}\big|_{T, \mu_{j \neq i}}$. The respective chemical potentials can be expressed through bulk concentrations from utilizing the electroneutrality condition in the bulk. These equations have been relegated to the SI, but the reader can readily obtain them from Eq. \eqref{eq:pressure}.

Taking the functional derivative of the grand canonical potential, Eq. \eqref{eq:BSK}, with respect to the electrostatic potential yields a modified 4th order BSK equation,
\begin{align}
     &\varepsilon_0 \varepsilon_s\left(1-l_c^2\frac{d^2}{dx^2}\right)\phi'' +\frac{d}{dx} \left(\frac{c_w \beta p_w^2\mathcal{L}(\beta p_w |\phi'|)}{\beta p_w|\phi'|}\phi'\right)=-e\left(c_+-c_- \right)
    \label{eq:mod_BSK}
\end{align}
where $\mathcal{L}(u)=\coth{(u)}-1/u$ is the Langevin function. Eq. \eqref{eq:mod_BSK} is solved here with the following 4 boundary conditions: $D|_{x=0} = \sigma$, where $\sigma$ is the surface charge density on the electrode; $\frac{d^3\phi}{dx^3}\Bigr|_{x=0} = 0$; $D|_{x=\infty} = 0$; and $\frac{d^3\phi}{dx^3}\Bigr|_{x=\infty} = 0$. 
\par 

The model parameters ($\xi_+$, $\xi_-$, $\gamma$, $\gamma_w$, $p_w$, $\varepsilon_s$, $l_c$) were extracted almost entirely from MD simulations directly (for details see the SI). In order to validate our theory, we compare its results with those obtained from MD simulations. We perform constant charge simulations for a set of electrode charges between -0.2 C/m$^2$ to +0.2 C/m$^2$, corresponding to potential drops between -2.43 V to 2.97 V with a potential of zero charge of -0.63 V. For the electrode, we use Lennard-Jones spheres arranged in an fcc lattice (100) with a lattice parameter of gold. Note that this electrode material effectively serves as a placeholder for electrode charge in the simulation; there are no species-specific electrolyte interactions with the electrode. Our goal in this work was not to study electrode-specific effects, but rather the fundamental properties of the WiSE EDL. Further details of the simulations are given in the SI. 
\par
\begin{figure*}[!hbt]
\centering
\begin{subfigure}[B]{0.49\textwidth}
    \includegraphics[clip, trim=3.5cm 8cm 3cm 8cm, width=1\textwidth]{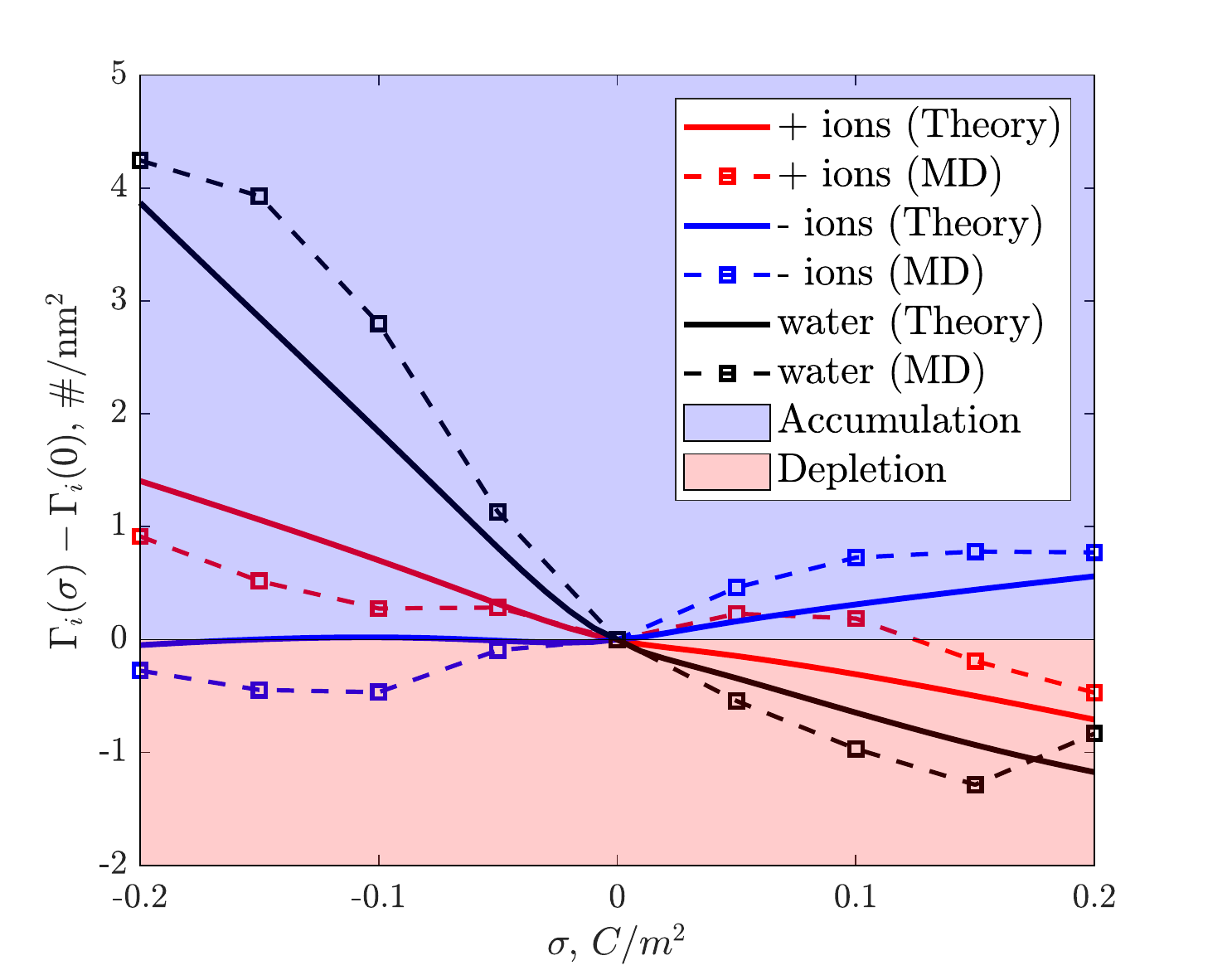}
    \caption{}
\end{subfigure}
\begin{subfigure}[B]{0.49\textwidth}
        \includegraphics[clip, trim=3.5cm 8cm 3cm 8cm, width=1\textwidth]{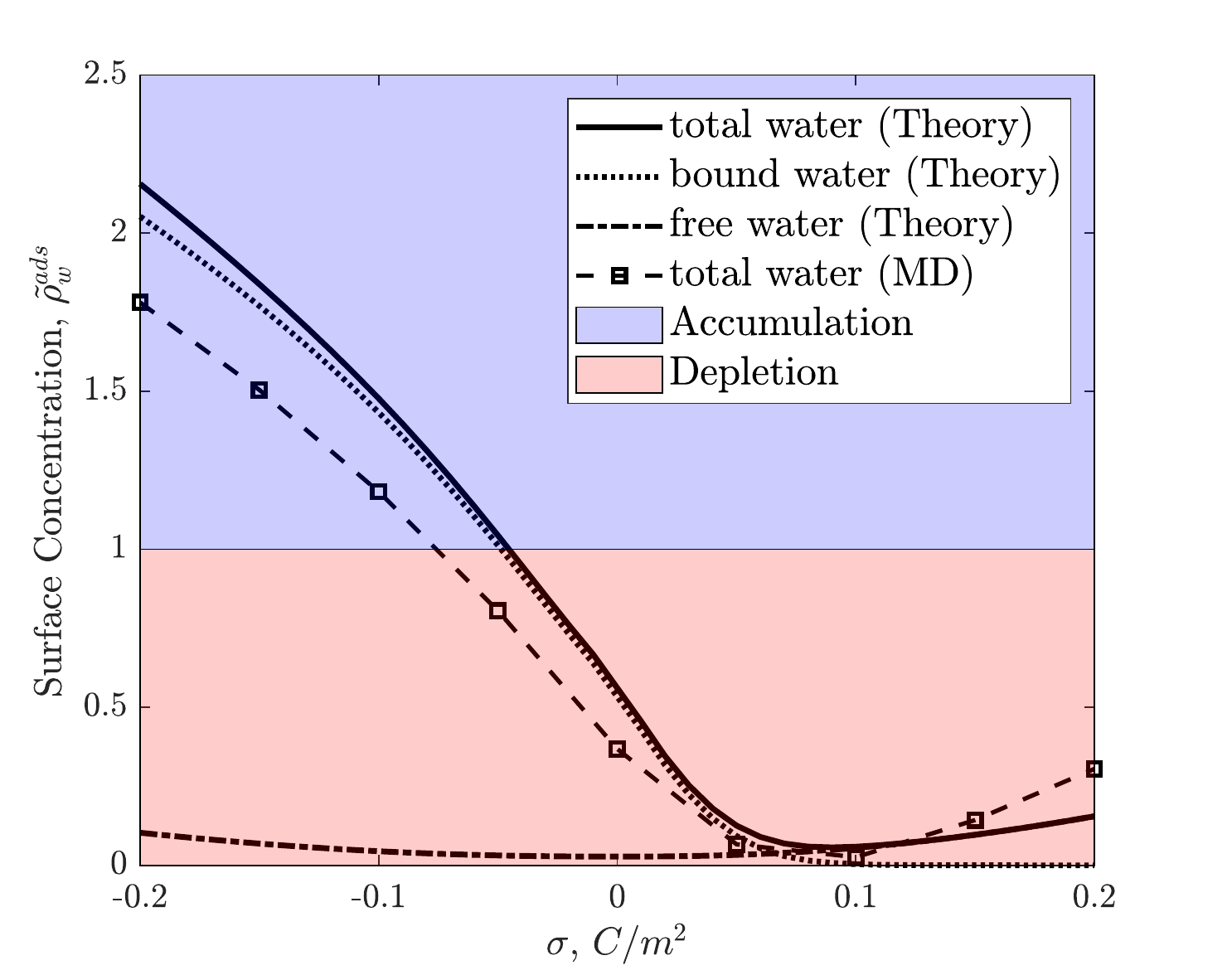}
        \caption{}
\end{subfigure}
\begin{subfigure}[B]{0.49\textwidth}
        \includegraphics[clip, trim=3.5cm 8cm 3cm 8cm, width=1\textwidth]{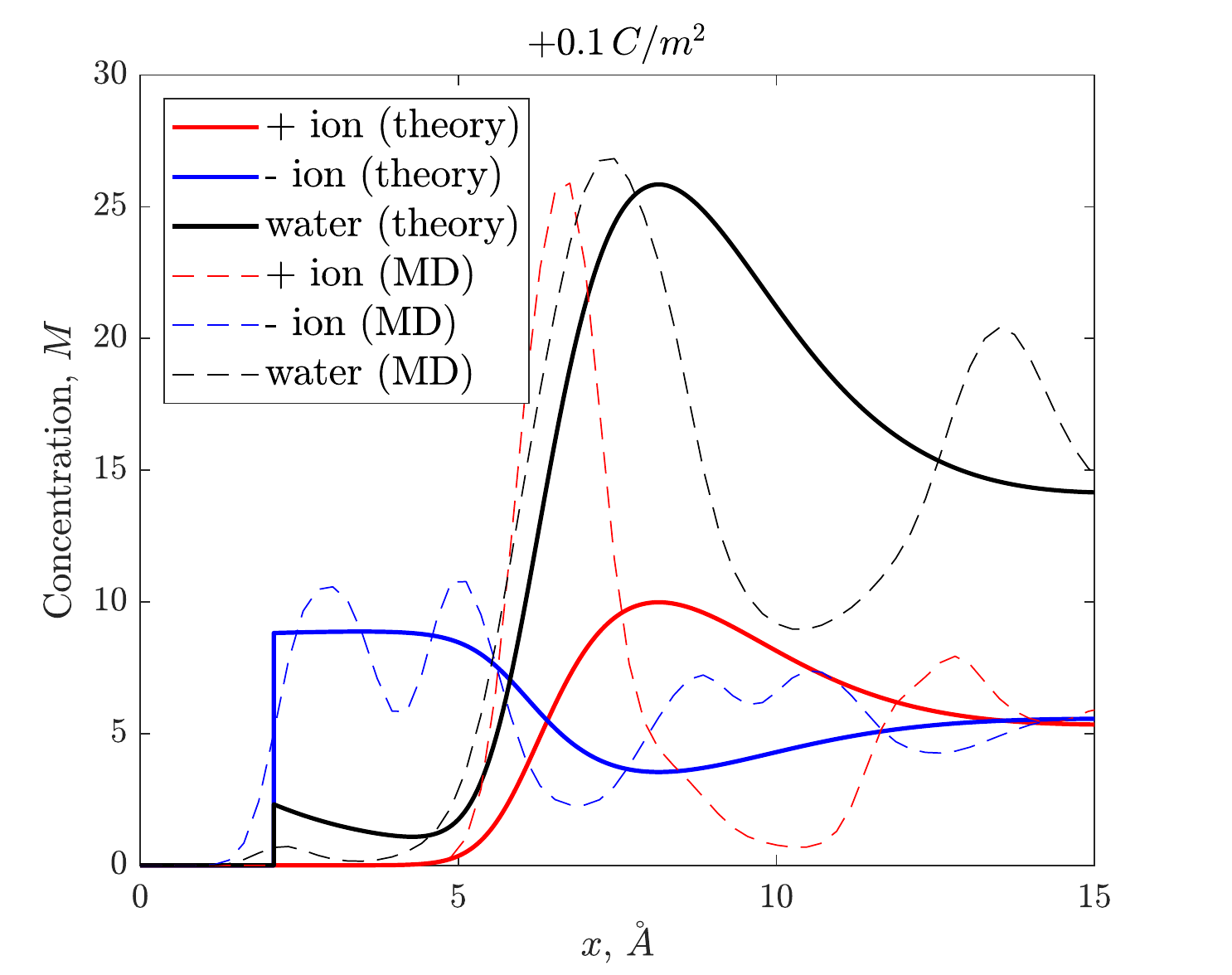}
        \caption{}
\end{subfigure}
\begin{subfigure}[B]{0.49\textwidth}
    \includegraphics[clip, trim=3.5cm 8cm 3cm 8cm, width=1\textwidth]{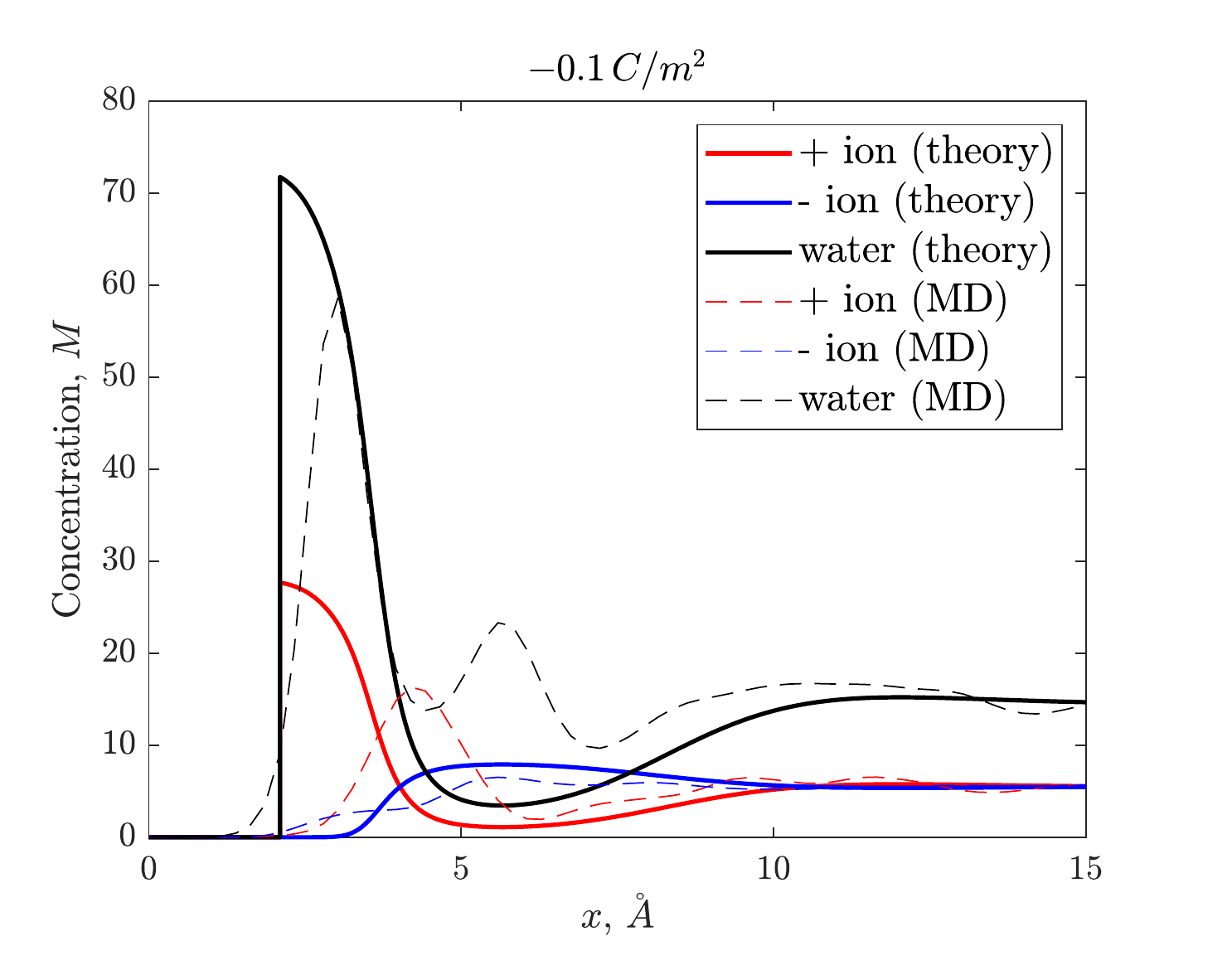}
    \caption{}
\end{subfigure}

\caption{(A) Surface excess concentration isotherms for all species obtained via Eq.\eqref{eq:Gamma} for both the theory and MD simulations. (B) The interfacial concentration of water integrates within 5 $\AA$ of the electrode surface. 
The concentration profiles of species in 21m LiTFSI next surfaces with charge (C) $+0.1$ C/m$^2$ and (D) $-0.1$ C/m$^2$}
\label{fig:esorb_litfsi}
\end{figure*}

In fig. \ref{fig:esorb_litfsi}A, we show electrosorption isotherms--the surface excess concentrations as a function of applied surface charge--for water as well as the ionic species. Surface excess quantities are the thermodynamically relevant quantities used in interfacial conservation laws of microscopically diffuse interfaces \cite{hiemenz1997,chu2007} and are defined with the following formula 
\begin{align}
    \Gamma_i(\sigma) = \int_0^{\infty}dx(c_i(x,\sigma)-c_i^{\text{bulk}}) 
    \label{eq:Gamma}
\end{align}
\par
 In our simulations we observe that even when there is zero surface charge, the simulated WiSE establishes interfacial structure. Namely, the TFSI$^-$ ion will wet the electrodes surface. For simplicity, we have not included any non-electrostatic species-electrode interactions in our theory, although such physics has been incorporated in previous continuum models.\cite{BBKGK}. To facilitate the comparison between the theory and MD simulations, we instead plot the deviation of the surface excess concentration from its value at 0 charge, \emph{i.e.}, $\Gamma_i(\sigma)-\Gamma_i(0)$.
\par
As in previous molecular simulations of WiSEs \cite{Vatamanu2017}, our simulations display a large asymmetry in the electrosorption of water in LiTFSI. This can be rationalized by the presence of competing solvation and polarization forces acting on water molecules in the EDL. In all cases, as the surface becomes more charged, polarization forces pull dipolar water molecules towards regions of high electric field, i.e., toward the surface \cite{Feng2014}. Next to negative electrodes, lithium counterions accumulate, bringing with them water of hydration, enriching the EDL with water molecules. However, next to positive electrodes, lithium ions will be depleted from the surface, and thus the bound water molecules will be depleted as well. 

\par
As can be seen in fig. \ref{fig:esorb_litfsi}, our theory is able to capture these trends in water electrosorption quite well. In 21m LiTFSI, we find that about 94\% of water molecules are bound to lithium, and thus most of the water in solution will follow the lithium electrosorption isotherm. The small amount of free water remaining will be subject to polarization forces, and will adsorb primarily in the first molecular layer next to the electrode.
\par

\par
Although the surface excess concentrations give us a picture of the entire EDL, the chemistry that governs the ESW in WiSE is the region within a few angstroms of the electrode surface. Therefore, it is also useful to understand how this interfacial region evolves as a function of electrode bias. In fig. \ref{fig:esorb_litfsi}B, we plot the dimensionless average concentration of water within 5 $\AA$ of the surface, $\tilde{\rho}^{ads}_w(L = 5 \AA,\sigma) = \int_{0}^{L} dxc_w(x,\sigma)/Lc^{\text{bulk}}_w$. When computing the theoretical curves, we shift the concentration profiles by the distance of closest approach and also separately plot the bound and free contributions. Our model captures the behavior of water sorption in this region remarkably well. Furthermore, we clearly see the theory predicts the large depletion of water next to positively biased electrodes, which may be responsible, in part, for the suppression of oxygen evolution and the expanded ESW in LiTFSI-based WiSEs.
\par
In figs. \ref{fig:esorb_litfsi}C and D we focus on the EDL profiles of all species for $\pm 0.1$ C/m$^2$. Clearly, the theory is capable of qualitatively resolving the interfacial profiles and molecular layering of the EDL in LiTFSI-based WiSE, observed in MD simulations. The fine structure of molecular layering is not captured in such a simple theory, but it actually recovers the general accumulation/depletion regions remarkably well. Another deficiency is that we assume the Li$^{+}$ and bound water act as a single, symmetric entity. As can be seen in the distributions, the Li$^{+}$ and bound water are orienting subject to electric field and neighbouring anions. Both of these features, fine structure and orientation of solvated Li$^{+}$, are inconsequential when calculating integrated quantities, which is why the theory is in good agreement with the performed simulations.
\par
\begin{figure}[!hbt]
    \centering
    \begin{subfigure}[B]{0.4\textwidth}
        \includegraphics[trim=5cm 8cm 5cm 8cm, width=1\textwidth]{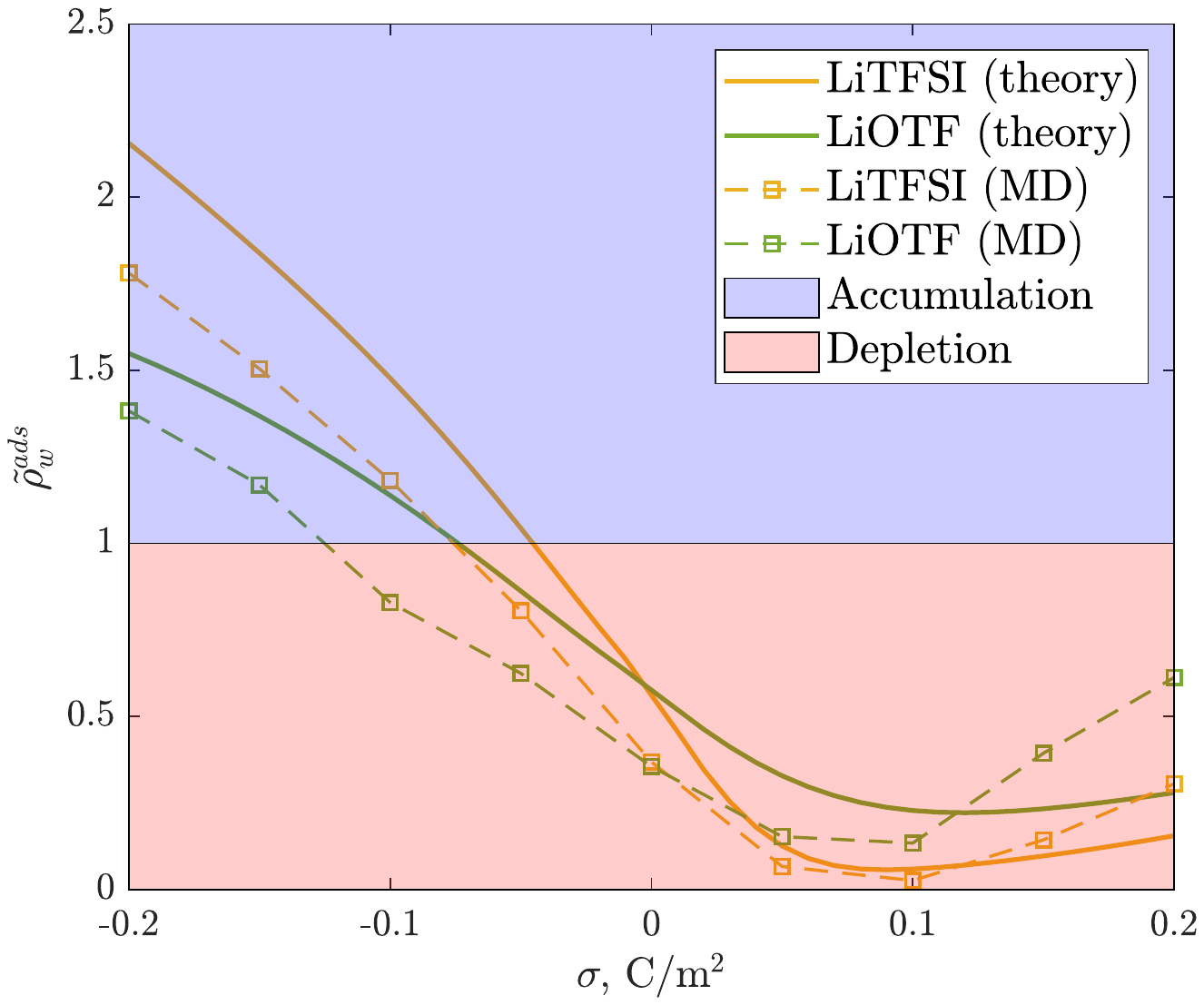}
        \caption{}
    \end{subfigure}
    \\
    \begin{subfigure}[B]{0.4\textwidth}
        \includegraphics[width=1\textwidth]{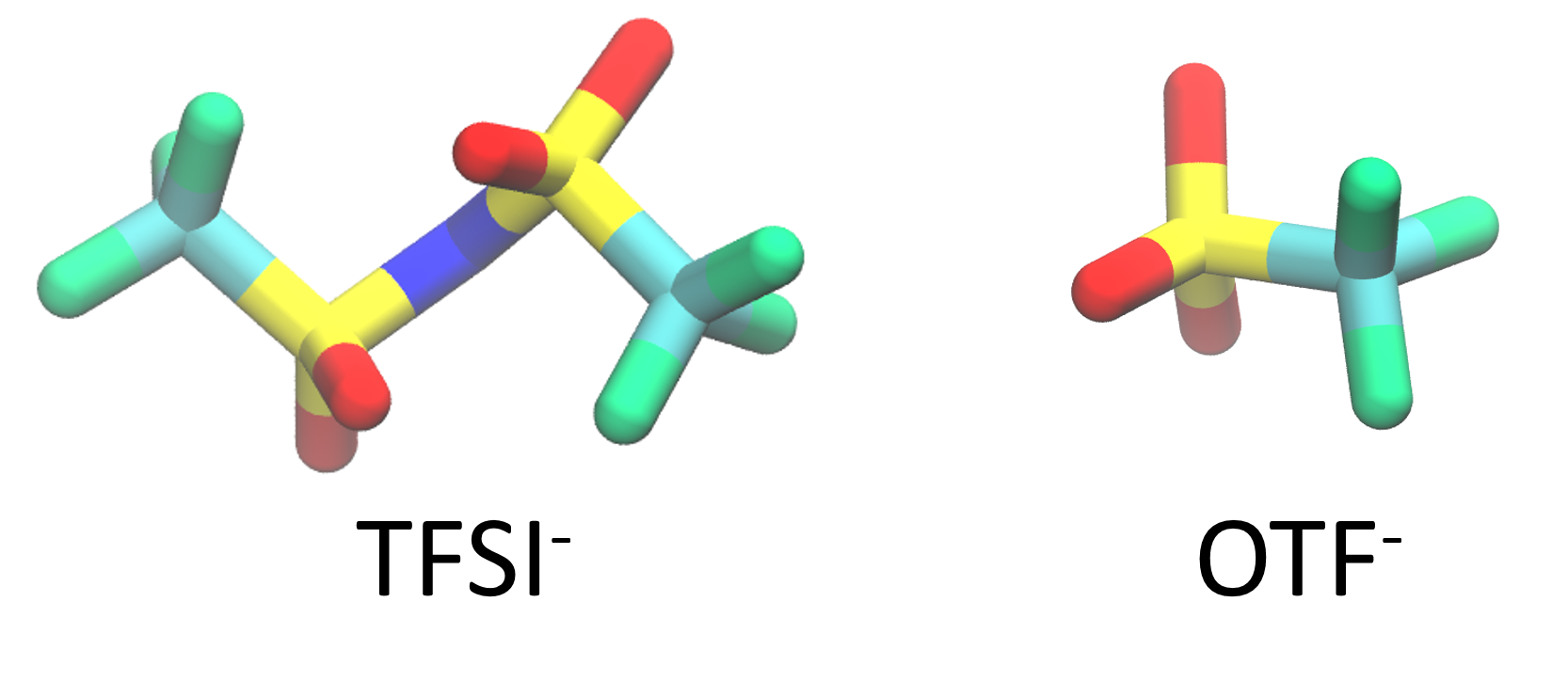}
        \caption{}
    \end{subfigure}
    \caption{A) A comparison of theoretical and simulated interfacial water sorption isotherms in 21m LiTFSI vs 21m LiOTF. The continuum theory is able to qualitatively capture the differences in the electrosorption curves, between two ostensibly similar salts. B) The chemical structures of TFSI$^-$ and OTF$^-$ ions are shown. }
    \label{fig:liotf_litfsi}
\end{figure}

Finally, we tested the ability of the model to capture salt-specific effects in electrosorption, by comparing two ostensibly similar salts, LiTFSI and lithium trifluoromethanesulfonate (LiOTF), at a concentration of 21m. The model is able to incorporate some salt-specific effects via the ion size, effective dipole moment, and fraction of free/bound water. In fig. \ref{fig:liotf_litfsi}, we plot the integrated interfacial water concentrations for LiTFSI and LiOTF obtained from theory and simulation. Understandably, the sign of asymmetry of water electrosorption is opposite to that found in ref. \citenum{Feng2014} where both of the ions were organic. There, water had a propensity to the anion, and there was more water in positively polarized EDL.
\par
The overall asymmetry of electrosorption is similar for both LiTFSI and LiOTF solutions. It can be seen, however, that at negative biases the accumulation of water is greater in LiTFSI than LiOTF. The reverse is true for positive biases, however. This phenomena is clearly captured by the theory, and the origin is the difference in the fraction of bound vs free water in LiOTF vs LiTFSI. In LiTFSI, 94\% of the water is bound, whereas, in LiOTF only 75\% of the water is bound. Thus, in the case of LiOTF, less water is attracted to the negative surface and repelled from the positive surface, because less water is bound to lithium than in the LiTFSI electrolyte. The origin of this phenomena is that there is a higher degree of ion pairing in LiOTF solutions, than in LiTFSI solutions. Thus, water molecules must compete with OTF$^-$ ions, in order to solvate Li$^+$. Although our theory does not explicitly treat these complex molecular interactions \cite{Chen2017}, we are able to model some of the effects these interactions have on EDL structure through the simple bound and free water partitioning. 
\par
In this Letter, we have proposed a simple continuum theory of WiSE, parameterized with MD simulations, which has the essential components necessary to reproduce the EDL structures observed in molecular simulations. These EDL structures play a critical role in the enhanced electrochemical stability of WiSEs observed  experimentally. Although similar mechanisms have been proposed before to explain the ESW enhancement,\cite{Vatamanu2017}, here we have gone a step further in understanding and quantifying the competing solvation and polarization forces acting on water molecules near electrofied interfaces in WiSEs. Although MD simulations can give us a useful prediction of the EDL structure of WiSEs\cite{Vatamanu2017}, the underlying physics is often difficult to infer from the complexity of the simulation. In contrast, the simplicity of our continuum model allows us to understand why certain EDL structures form and how to engineer WiSEs to yield more desirable EDL structures.

\section{Supporting Information}
\subsection{Molecular Dynamics Simulation Methodology}
In this study, we performed all-atom classical MD simulations using LAMMPS \cite{plimpton1995}. We performed a set of simulations for two different salts: LiTFSI and LiOTF. For each salt we first performed a series of bulk-like, fully periodic simulations at molal concentrations of 7m, 10m, 12m, 15m, and 21m. This yielded bulk densities, ion-water correlation functions, and data for fitting model parameters ($p_w$ and $\varepsilon_s$). Additionally, we performed MD simulations of the 21m solution in a nano-slit geometry with surface charges of $0,\pm 0.05,\pm 0.1, \pm 0.15,$ and $\pm 0.2$ C/m$^2$.
\par
\begin{figure*}[!hbt]

\centering
\begin{subfigure}[B]{2.8in}
    \includegraphics[clip,trim=2.5cm 3cm 2.5cm 5cm, width=1\textwidth]{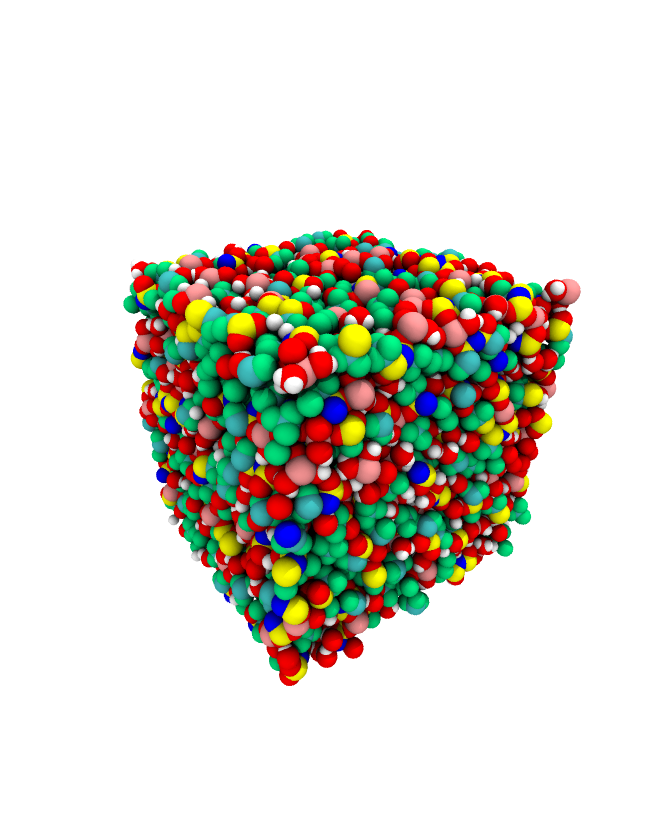}
    \caption{}
\end{subfigure}
\begin{subfigure}[B]{5in}
    \includegraphics[clip,trim=0cm 12cm 0cm 12cm, width=1\textwidth]{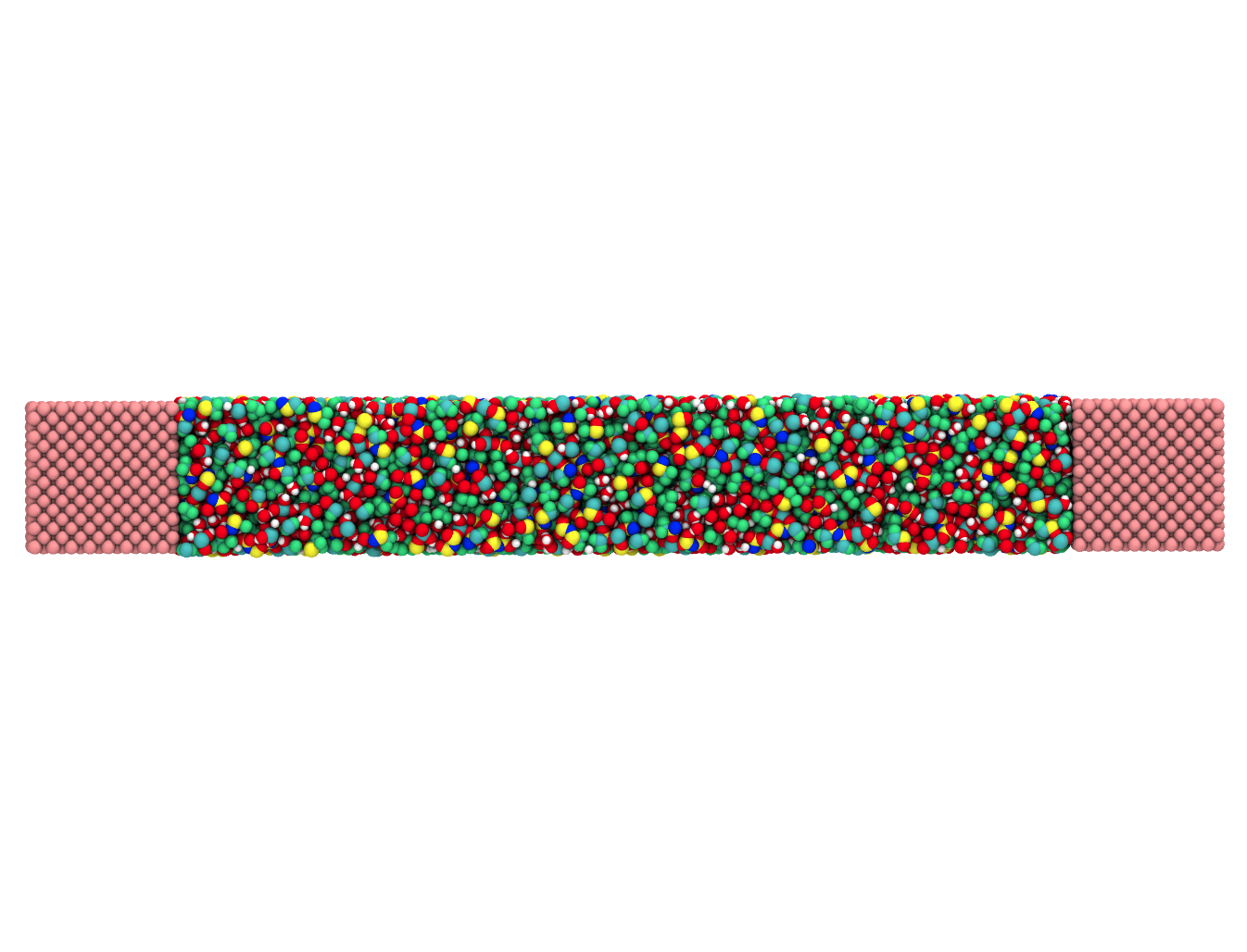}
    \caption{}
\end{subfigure}

\caption{Snapshots of molecular dynamics simulations in (A) periodic and (B) nano-slit geometries.}
\label{fig:MD_sims}
\end{figure*}

\emph{Simulation Details:} In the periodic geometries, as seen in fig. \ref{fig:MD_sims}A., we performed simulations containing 1000 water molecules and enough ion pairs for 7m, 10m, 12m, 15m, and 21m solutions. The simulations were performed at fixed temperature (300 K) and pressure (1 bar), with Nose-Hoover thermostat and barostat until the density of the fluid equilibrated (10 ns with 1 fs time steps). Next, we switched to constant volume simulation box, still with a fixed temperature (300 K) and Nose-Hoover thermostat, and equilibrate for an additional 6 ns. Finally, production runs were performed for an additional 6 ns. 
\par
In the nano-slit geometries, as seen in fig. \ref{fig:MD_sims}B., we simulated the system at constant volume and temperature, filling a 33x33x200 $\AA^3$ simulation box, with two 33x33x33 $\AA^3$ electrodes made up of Lennard Jones (LJ) spheres arranged in an fcc lattice (100) corresponding to gold. We refer to these LJ spheres as `gold' atoms from here on out, although the primary purpose is to hold an applied charge and serve as a hard boundary for the  electrolyte. The electrodes are arranged to sandwich the electrolyte fluid. For 21m LiTFSI (LiOTF), the box contained 707 (1096) ion pairs, 1873 (2904) water molecules, and 4096 (4096) `gold' atoms. Surface charges ($0,\pm 0.05,\pm 0.1, \pm 0.15,$ and $\pm 0.2$ C/m$^2$) were applied by placing partial charges on the first layer of `gold' atoms.  Equilibration runs of about 12 ns (1 fs time steps) were performed initially with no applied potential/charge. Then the surface charge was ramped up from zero, allowing for 12 ns of equilibration and 6 ns of production at each electrode surface charge.
\par
The initial configurations for all simulations were generated using the open-source software, PACKMOL \cite{martinez2009packmol}. All MD simulations were visualized using the open-source software, VMD \cite{HUMP96}.
\par
\emph{Force Field Details}: For all ionic species we employed the CL$\&$P force field, which was developed for ionic liquid simulations, with same functional form as the OPLSAA force field \cite{lopes2012}. Given the dense ionic nature of our systems, we expect the CL$\&$P force field to be appropriate for WiSEs. For water, we employed the spc/e force field. Interatomic interactions are determined using Lorentz-Berthelot mixing rules. Finally, for nano-slit simulations, we require force fields for the `gold' electrode. We did not explicitly model the dynamics of the electrode, omitting the need for a `gold'-`gold' force field. The `gold' interacts with the fluid mainly via coulomb interactions, as the surface layer of `gold' atoms are charged. We also include Lennard-Jones interactions, which were made to be the same no matter what atom is interacting with `gold' (LJ well depth: $\varepsilon = 0.001\text{eV}$, LJ well distance: $\sigma = 3 \AA$). We make the Lennard-Jones parameters constant for all species in order to emphasize the role of the electrolyte and obtain conclusions that are not specific to the electrode material. Long range electrostatic interactions were computed using the Particle-Particle Particle-Mesh (PPPM) solver (with a cut-off length of 12 $\AA$), which maps particle charge to a 3D mesh for the periodic simulations and a 2D mesh in the transverse direction for the nano-slit simulation\cite{hockney1988}. 

\newpage
\subsection{Electrostatic Energy:}
We start by writing the displacement field:
\begin{align}
    \textbf{D} = \varepsilon_0 \textbf{E}+ \textbf{P}_{\text{el}}+\textbf{P}_{\text{dip}}+\textbf{P}_{\text{BSK}}
    \label{eq::D}
\end{align}
where $\textbf{E}=-\nabla\phi$ is the electric field and $\phi$ is the electrostatic potential; $\varepsilon_0$ is vacuum permittivity; $\textbf{P}_{\text{el}}$ is the local background polarization field of the electrolyte, with major contributions coming from the electronic polarizability of molecular species which is assumed to be constant; $\textbf{P}_{\text{dip}}$ is the local dipolar polarization field and is assumed to be a strong function of the concentration of dipoles, $c_w$; and $\textbf{P}_{\text{BSK}}$ is the local polarization field due to the orientation of correlated ions, as described by the Bazant-Storey-Kornyshev (BSK) theory \cite{Bazant2011}. 
%
%
\par
Each of the contributions to the polarization field can be written as
\begin{align}
    \textbf{P}_{\text{el}}=\varepsilon_0 (\varepsilon_{\text{s}}-1) \textbf{E}
    \label{eq:Pinfty}
\end{align}
\begin{align}
    \textbf{P}_{\text{dip}}=c_w \langle p_w \rangle=c_w p_w \mathcal{L}(\beta p_w \textbf{E})
    \label{eq:Pdipole}
\end{align}
\begin{align}
    \textbf{P}_{\text{BSK}} = -\varepsilon_0\varepsilon_s l_c^2 \nabla^{2}\textbf{E}
    \label{eq:Pbsk}
\end{align}
where $p_w$ is the effective dipole moment of free water molecules; $\beta=1/k_BT$ is inverse thermal energy; $\mathcal{L}(u) = \coth{u} - 1/u$ is the Langevin function; and $l_c$ is the ionic correlation length \cite{Bazant2011}. In reality, $\textbf{P}_{\text{el}}$ is a function of the local concentrations of all the species in the fluid; this contribution turns out to be roughly constant here. Therefore, in practice, we define $\varepsilon_{\text{s}}$ to be the permittivity of the WiSE when there is no free water in the fluid, which is something that we will obtain from fitting MD simulations. 

It is convenient to write the displacement field in terms of linear and non-linear contributions:
\begin{align}
    \textbf{D}=\textbf{P}_\text{dip}+\varepsilon_0\varepsilon_s(1-l_c^2\nabla^2)\textbf{E}
\end{align}

The differential electrostatic energy, $\delta W_e$ for a dielectric medium of volume, $V$, is given by 
\begin{align}
    \delta W_e = \int_V d\textbf{x}\left[\textbf{E} \cdot \delta \textbf{D}\right]     
    \label{eq:delW_e}
\end{align}
where we may write the variation in the displacement field as the following
\begin{align}
    \delta \textbf{D} &= \frac{\delta \textbf{P}_\text{dip}}{\delta \textbf{E}} \delta \textbf{E}+\varepsilon_0\varepsilon_s\delta\textbf{E}-\varepsilon_0\varepsilon_s l_c^2 \nabla\delta(\nabla \cdot \textbf{E})
\end{align}
where we have used the identity $\nabla\cdot\nabla\textbf{E}=\nabla \nabla\cdot \textbf{E}$, when $\nabla\times\textbf{E}=0$. We then plug this result into Eq. \eqref{eq:delW_e}
\begin{align}
    \delta W_e = \int_V d\textbf{x}\left[ \frac{\delta \textbf{P}_\text{dip}}{\delta \textbf{E}}\textbf{E} \cdot \delta \textbf{E}+\varepsilon_0\varepsilon_s \textbf{E}\cdot \delta \textbf{E} -\varepsilon_0\varepsilon_s l_c^2 \textbf{E}\cdot \nabla\delta(\nabla \cdot \textbf{E})\right]
\end{align}
We may then use the vector identity:
\begin{align}
\textbf{E}\cdot\nabla\delta(\nabla\cdot\textbf{E})=\nabla\cdot(\textbf{E}\delta(\nabla\cdot\textbf{E}))-\nabla\cdot\textbf{E}\delta(\nabla\cdot\textbf{E})
\label{eq:identity}
\end{align}
and employed the divergence theorem to transform the first term of Eq. \eqref{eq:identity} into a surface term, at which the electric field vanishes. Thus 
\begin{align}
    \delta W_e = \int_V d\textbf{x}\left[ \frac{\delta \textbf{P}_\text{dip}}{\delta \textbf{E}}\textbf{E} \cdot \delta \textbf{E}+\varepsilon_0\varepsilon_s \textbf{E}\cdot \delta \textbf{E} +\varepsilon_0\varepsilon_s l_c^2 \nabla \cdot \textbf{E}\cdot \delta(\nabla \cdot \textbf{E})\right]
\end{align}

Integrating $\delta W_e$ from 0 to $W_e$, plugging in the definition, $\textbf{E}=-\nabla \phi$, and enforcing Poisson's equation, $\nabla \cdot \textbf{D}=\rho$, with a Lagrange multiplier, $\lambda$, obtains:

\begin{align}
    W_e &= \int_V d\textbf{x}\left[ \frac{c_w}{\beta}\left(\beta p_w \nabla \phi \cdot \mathcal{L}(\beta p_w \nabla \phi)-\ln \left( \frac{\sinh(\beta p_w \nabla \phi)}{\beta p_w \nabla \phi} \right)\right)\right] \nonumber \\ 
    &+\int_V d\textbf{x}\left[\frac{\varepsilon_s}{2}\left(|\nabla \phi|^2+l_c^2|\nabla^2 \phi|^2\right)\right] \nonumber \\
    &+\int_V d\textbf{x}\left[ \lambda\left(\rho- \nabla \cdot \textbf{D} \right) \right] 
\end{align}
Substituting in Eq. \eqref{eq::D} and integrating by parts shows that $\lambda=\phi$. Hence the electrostatic energy of the WiSE is given by:
\begin{align}%
W_e =& \int_V w_e d\mathbf{x} \\
=&\int_V d\mathbf{x}\bigg[\rho \phi -\frac{\varepsilon_0\varepsilon_s}{2}\left(|\nabla \phi|^2+l_c^2|\nabla^2 \phi|^2\right)
- c_w k_B T \ln \left(\frac{\sinh \left( \beta p_w |\nabla \phi| \right)}{\beta p_w |\nabla \phi|} \right)\bigg]
\label{eq:we_final}
\end{align}

\subsection{Pressure Function:} 

\noindent Here we outline a heuristic derivation of the pressure function for a Langmuir model. This allows the form of the pressure function to be derived, without going into too much detail.

The grand canonical partition function for a 3 component system can be seen as:

\begin{equation}
\Xi = \sum_{n_{+} = 0}^{\infty}\sum_{n_{-} = 0}^{\infty}\sum_{n_{w} = 0}^{\infty}e^{\beta\mu_{+}n_{+}}e^{\beta\mu_{-}n_{-}}e^{\beta\mu_{w}n_{0}}Q(n_{+})Q(n_{-})Q(n_{w})
\end{equation}

\noindent where $Q(n_{j})$ corresponds to the canonical partition function of each species, $j$, as seen by

\begin{equation}
Q(n_{j}) = \Omega_{j}e^{-\beta \epsilon_{j}n_{j}}
\end{equation}

\noindent where the energy is given by $E_{j} = \epsilon_{j}n_{j}$.

In the model of Han \textit{et al.}\cite{Han2014} it was assumed that the lattice is subsequently filled with each species. Hence, the configurational entropy is written as if each component, of which there are $n_{j}$ individual elements, resides on its own `individual lattice',

\begin{equation}
\Omega_{j} = \dfrac{N_{t}!}{n_{j}!(N_{t} - n_{j})!}
\end{equation}

Hence, we have for the partition function

\begin{equation}
\Xi = \sum_{n_{w} = 0}^{N}\Omega_{w}e^{\beta\tilde{\mu}_{w}n_{w}}\cdot \sum_{n_{-} = 0}^{(N - n_{w})/\xi_{-}}\Omega_{-}e^{\beta\tilde{\mu}_{-}n_{-}}\cdot \sum_{n_{+} = 0}^{(N - n_{w} - \xi_{-}n_{-})/\xi_{+}}\Omega_{+}e^{\beta\tilde{\mu}_{+}n_{+}}
\end{equation}

\noindent where $\xi_{-} = v_{-}/v_{w}$ and $\xi_{+} = v_{+}/v_{w}$, and $\tilde{\mu}_{j}=\mu_j-\epsilon_j$ has been introduced for convenience.

The sums are evaluated progressively because of the assumption that the lattice is subsequently filled with each component. Using the binomial theorem, we obtain after evaluating the sums





\begin{equation}
\Xi = \Big\{[(1 + e^{\beta\tilde{\mu}_{+}})^{\xi_{-}/\xi_{+}} + e^{\beta\tilde{\mu}_{-}}]^{1/\xi_{-}} + e^{\beta\tilde{\mu}_{w}}\Big\}^{N}
\end{equation}

\noindent In the limit of all components of equal size, we recover the well-known symmetric 3 component Langmuir partition function:

\begin{equation}
\Xi = \Big\{1 + e^{\beta\tilde{\mu}_{+}} + e^{\beta\tilde{\mu}_{-}} + e^{\beta\tilde{\mu}_{w}}\Big\}^{N}
\end{equation}

In the grand canonical ensemble the partition function is related to the pressure, $p$, through,

\begin{equation}
pV = k_{B}T\ln\Xi
\end{equation}

\noindent Hence, we have for the pressure function,

\begin{equation}
p = \dfrac{k_{B}T}{v_{w}}\ln \{ [(1 +e^{\beta\tilde{\mu}_{+}})^{\xi_{-}/\xi_{+}} + e^{\beta\tilde{\mu}_{-}}]^{1/\xi_{-}} + e^{\beta\tilde{\mu}_{w}}\}
\end{equation}

\noindent To obtain the pressure function in the main text, we bring a constant out of the chemical potentials:

\begin{equation}
p = \dfrac{k_{B}T}{v_{w}}\ln \{ [(1 + \xi_{+}e^{\beta\tilde{\mu}_{+}})^{\xi_{-}/\xi_{+}} + \xi_{-} e^{\beta\tilde{\mu}_{-}}]^{1/\xi_{-}} + e^{\beta\tilde{\mu}_{w}}\}
\end{equation}

It should be evident that this derivation, which is a generalization of the special case of ref. \citenum{Han2014}, can be further generalized to any number of components of any size. 





\subsection{Concentrations:}


The concentrations can be obtained from 

\begin{align}
c_{j} = \frac{\partial p}{\partial \mu_{j}}\bigg|_{T, \mu_{i \neq j}}
\end{align}

\noindent Taking these derivatives yields

\begin{align}
     \tilde{c}_{w} = \frac{2}{\gamma}\dfrac{e^{\beta(\mu_{w}+\Psi)}}{[(1 + \xi_{+}e^{\beta(\mu_{+}-e\phi)})^{\xi_{-}/\xi_{+}} + \xi_{-} e^{\beta(\mu_{-}+e\phi)}]^{1/\xi_{-}} + e^{\beta(\mu_{w}+\Psi)}}
     \label{eq:cw}
\end{align}
\begin{align}
\tilde{c}_{-} = \frac{2}{\gamma}\dfrac{e^{\beta(\mu_{-}+e\phi)}[(1 + \xi_{+}e^{\beta(\mu_{+}-e\phi)})^{\xi_{-}/\xi_{+}} + \xi_{-} e^{\beta(\mu_{-}+e\phi)}]^{1/\xi_{-} - 1}}{[(1 + \xi_{+}e^{\beta(\mu_{+}-e\phi)})^{\xi_{-}/\xi_{+}} + \xi_{-} e^{\beta(\mu_{-}+e\phi)}]^{1/\xi_{-}} + e^{\beta(\mu_{w}+\Psi)}}
\label{eq:c-}
\end{align}
\begin{align}
\tilde{c}_{+} = \frac{2}{\gamma}\dfrac{e^{\beta(\mu_{+}-e\phi)}(1 + \xi_{+} e^{\beta(\mu_{+}-e\phi)})^{\xi_{-}/\xi_{+} - 1}[(1 + \xi_{+}e^{\beta(\mu_{+}-e\phi)})^{\xi_{-}/\xi_{+}} + \xi_{-} e^{\beta(\mu_{-}+e\phi)}]^{1/\xi_{-} - 1}}{[(1 + \xi_{+}e^{\beta(\mu_{+}-e\phi)})^{\xi_{-}/\xi_{+}} + \xi_{-} e^{\beta(\mu_{-}+e\phi)}]^{1/\xi_{-}} + e^{\beta(\mu_{w}+\Psi)}}
\label{eq:c+}
\end{align}
where $\gamma = 2c^{\text{bulk}}_\pm v_w$ is the packing parameter for ions in the bulk, and $\tilde{c_i}=c_i/c^{\text{bulk}}_\pm$ are non-dimensionalized concentrations. Note that the packing parameter is not equivalent to the typically defined compacity\cite{Kornyshev2007}, since we have an asymmetric lattice here.
\par 
In the limit of $\Psi \gg |e\phi|$ (see main text for definition of the function), $c_w \rightarrow 1/v_w$, while ions are depleted $c_\pm \rightarrow 0$. When $e\phi \gg \Psi > -e\phi$, the anions saturate, $c_- \rightarrow 1/\xi_{-}v_w = 1/v_-$, with all other concentrations tending to zero, $c_+,w \rightarrow 0$. For $-e\phi \gg \Psi > e\phi$, the cation concentration reaches a maximum, $c_+ \rightarrow 1/\xi_{+}v_w = 1/v_+$, with $c_-,w \rightarrow 0$. 
\par
The chemical potentials in \eqref{eq:cw}-\eqref{eq:c+} are determined by the concentrations of species in the bulk reservoir, where there are no electrostatic fields. Solving the system of algebraic equations, Eqs. \eqref{eq:cw}-\eqref{eq:c+}, under these conditions yields:
\begin{align}
\beta\mu_{w} = \ln\Bigg[\dfrac{\gamma_{w}}{1 - \gamma_{w}} \Bigg] + \dfrac{1}{\xi_{-}}\ln\Bigg[\dfrac{1 - \gamma_{w}}{1 - \gamma_{w} - \xi_{-}\gamma/2}\Bigg] + \dfrac{1}{\xi_{+}}\ln\Bigg[\dfrac{1 - \gamma_{w} - \xi_{-}\gamma/2}{1 - \gamma_{w} - \xi_{-}\gamma/2 - \xi_{+}\gamma/2} \Bigg] 
\end{align}
\begin{align}
\beta\mu_{-} = \ln\Bigg[\dfrac{\gamma/2}{1 - \gamma_{w} - \xi_{-}\gamma/2} \Bigg] + \dfrac{1}{\xi_{+}}\ln\Bigg[\dfrac{1 - \gamma_{w} - \xi_{-}\gamma/2}{1 - \gamma_{w} - \xi_{-}\gamma/2 - \xi_{+}\gamma/2} \Bigg] 
\end{align}
\begin{align}
\beta\mu_{+} = \ln\Bigg[\dfrac{\gamma/2}{1 - \gamma_{w} - \xi_{-}\gamma/2 - \xi_{+}\gamma/2} \Bigg]
\end{align}
where $\gamma_w = c^{\text{bulk}}_w v_w $ is the packing parameter free water molecules in the bulk. Note that $\xi_{-}\gamma/2$ is the comapcity of anions and $\xi_{+}\gamma/2$ is the compacity of cations, as they would be defined in a symmetric lattice-gas \cite{Kornyshev2007}.

\subsection{Model Parameterization:}
\label{sec:param}
Our theory is parameterized from molecular dynamics simulations of 21m aqueous solutions of LiTFSI and LiOTF in fully periodic geometries, representing the bulk-like fluid. The model developed in the main text and Appendix requires the specification of 7 parameters: $\xi_+$, $\xi_-$, $\gamma$, $\gamma_w$, $p_w$, $\varepsilon_s$, and $l_c$. Here, we discuss our procedure for determining these parameters. 
\par
\begin{figure}[!hbt]
    \centering
    \includegraphics[clip, trim=3cm 8cm 3cm 8cm, width=0.5\textwidth]{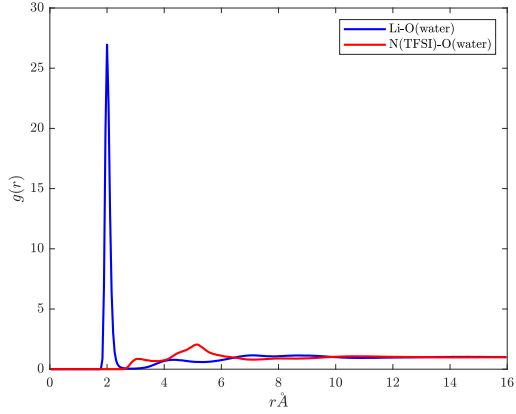}
    \caption{The Li-O(water) and N(TFSI)-O(water) radial distribution functions are plotted to contrast the asymmetry in ion solvation between anions and cations. The correlation functions serve in determining the contribution of ion solvation in the ion size parameters, $xi_+$ and $\xi_-$.}
    \label{fig:rdfs}
\end{figure}

In the main text we explained that $\xi_+$ had a contribution for solvating water molecules. On the other hand, we made the assumption that $\xi_-$ is only related to the molecular volume of the `naked' TFSI$^-$ ion. It can be seen in fig. \thefigure{fig:rdfs}, that cations molecules are much more vigorously solvated than anions. This observation serves as the justification for our assumption that no water is bound to anions.

\par

We determined the bulk volumetric filling fraction, $\Phi_B$, of the fluid, by fitting it to reproduce the bulk molar concentrations of species observed in MD simulations. For LiTFSI, we found a bulk filling fraction of 0.81. The bulk filling fraction is related to $\gamma$ and $\gamma_w$ in the following way:
\begin{align}
    \Phi_B = \gamma_w+\frac{\gamma}{2}(\xi_++\xi_-)
\end{align}
where $\gamma$ and $\gamma_w$ are related via the molality of the solution, and the number of bound water molecules. We write the equation for the ratio of ion pairs to water molecules:
\begin{align}
    \frac{1~\text{mol water}}{0.018~\text{kg}}\times \frac{1~\text{kg}}{21~\text{mol ion}}=2.65  \frac{\text{mol water}}{\text{mol ion}} 
\end{align}
We may then write 
\begin{align*}
    2.65 =\frac{\gamma_w+n_w^{\text{bound}}\gamma/2}{\gamma/2}
\end{align*}
where $n_w^{\text{bound}}$ is the average number of bound water molecules to the cation. Thus, upon determining $\Phi_B$, we specify both $\gamma$ and $\gamma_w$.  
\par
We now determine the remaining parameters, $p_w$ and $\varepsilon_s$, from a set of MD simulations with various water concentrations (7m-21m salt concentration) and computing the bulk dielectric constant, using the well-known Kirkwood formula: 
\begin{align}
    \varepsilon_{MD}=1 + \frac{4 \pi}{3 V k_B T}\left( \ave{|\textbf{M}|}^2-\ave{|\textbf{M}|^2} \right)
\end{align}
where $\textbf{M}$ is the total instantaneous dipole moment of the fluid during a snapshot of the MD simulation. 
\par
\begin{figure}[!hbt]
    \centering
    \includegraphics[clip, trim=3cm 8cm 3cm 8cm, width=0.5\textwidth]{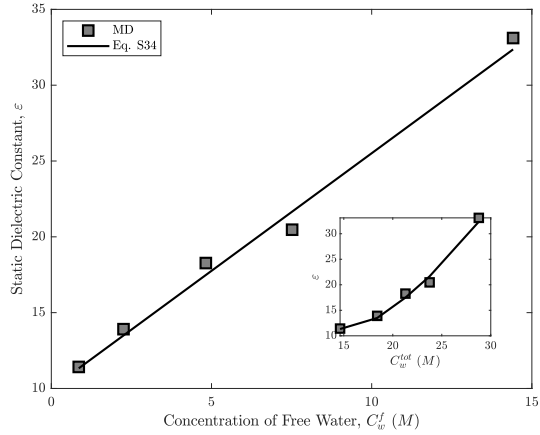}
    \caption{The static dielectric constant computed from MD simulations, as well as with Eq. \eqref{eq:lin_eps} is plotted as a function of \emph{free} water concentration. Inset: The dielectric constants are plotted as a function of \emph{total} water concentration. }
    \label{fig:permittivity}
\end{figure}

The model described in the previous section yields the following formula for the permittivity operator:
\begin{align}
    \hat{\varepsilon}=\frac{c_w \beta p_w^2}{\varepsilon_0} \frac{\mathcal{L}(\beta p_w \textbf{E})}{\beta p_w \textbf{E}}+\varepsilon_{\text{s}}\left( 1 - l_c^2 \nabla^{2}\right)
\end{align}
While parameterizing, the simulations were performed in the absence of an external electric field and averaged over the entire simulation box, so we may take limit of $\hat{\varepsilon}$ as $|\textbf{E}|\rightarrow0$ and the limit in which the gradient operators vanish (long wavelength limit):
\begin{align}
    \hat{\varepsilon} \rightarrow \varepsilon=\varepsilon_s+c_w\frac{p_w^2}{3\varepsilon_0k_BT}
    \label{eq:lin_eps}
\end{align}
We fit Eq. \eqref{eq:lin_eps} to the dielectric constants computed from MD in order to determine $p_w$ and $\varepsilon_s$, as shown in fig. \ref{fig:permittivity}. 
For LiTFSI, we obtained values of $\varepsilon_s=10.1$ and $p_w = 2.7 p_w^0$, where $p_w^0=1.85 ~\text{Debye}$ is the dipole moment of a water molecule in vacuum. 
\par
The last parameter we must specify is $l_c$, the correlation length, which is perhaps the least well determined parameter in the model. As a modification to Poisson-Fermi theory, $l_c$ was first introduced by BSK as a parameter which controls the magnitude of electrostatic correlations between charges in an IL \cite{Bazant2011}. These electrostatic correlations lead to the "overscreening" phenomena, which manifests as decaying oscillations in density profiles within the EDL. In ref. \citenum{Bazant2011}, $l_c$ was taken to be the ionic diameter.
\par
\begin{figure*}[!hbt]
\centering
\begin{subfigure}[B]{0.49\textwidth}
    \includegraphics[clip, trim=3.5cm 8cm 3cm 8cm, width=1\textwidth]{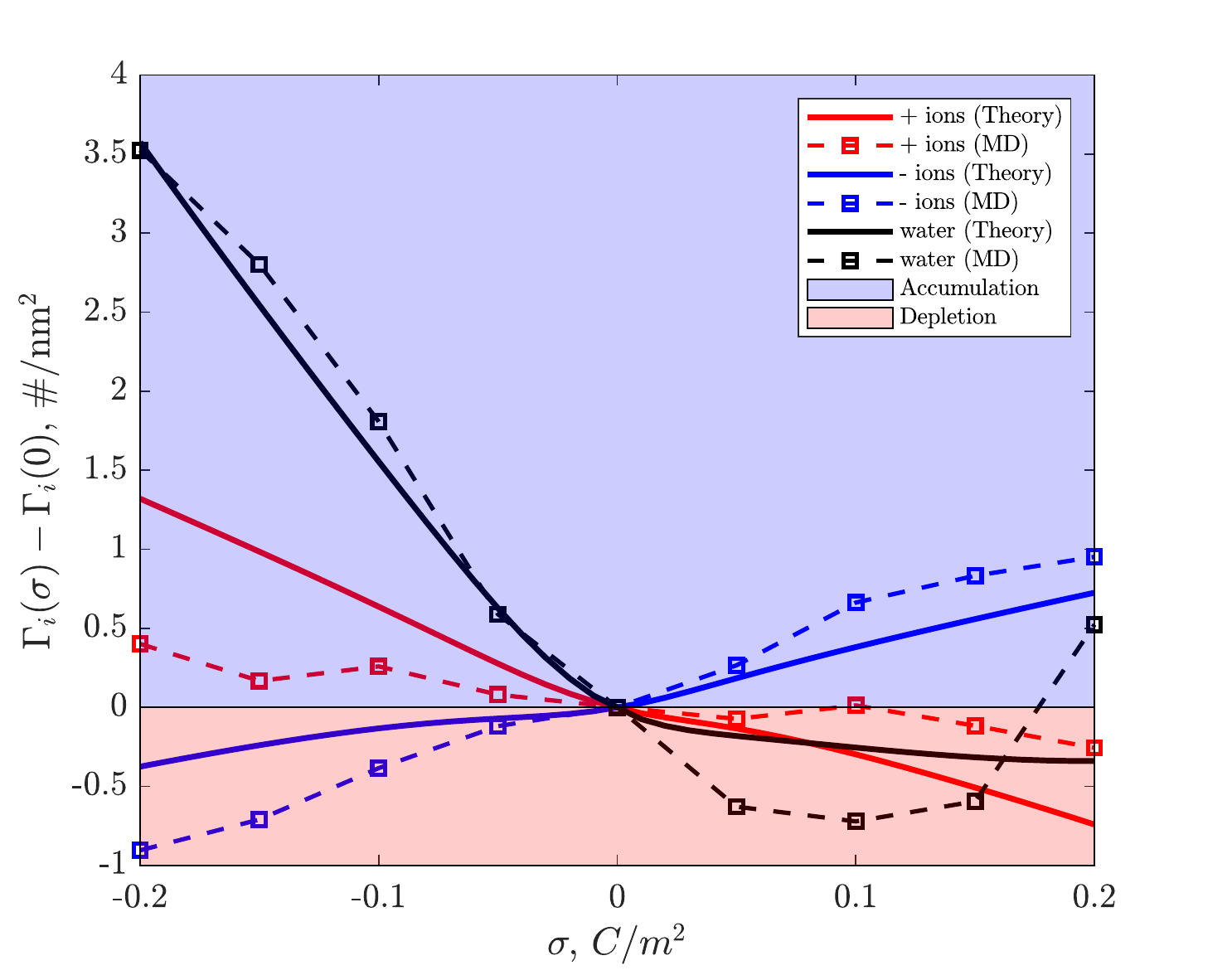}
\end{subfigure}
\begin{subfigure}[B]{0.49\textwidth}
    \includegraphics[clip, trim=3.5cm 8cm 3cm 8cm, width=1\textwidth]{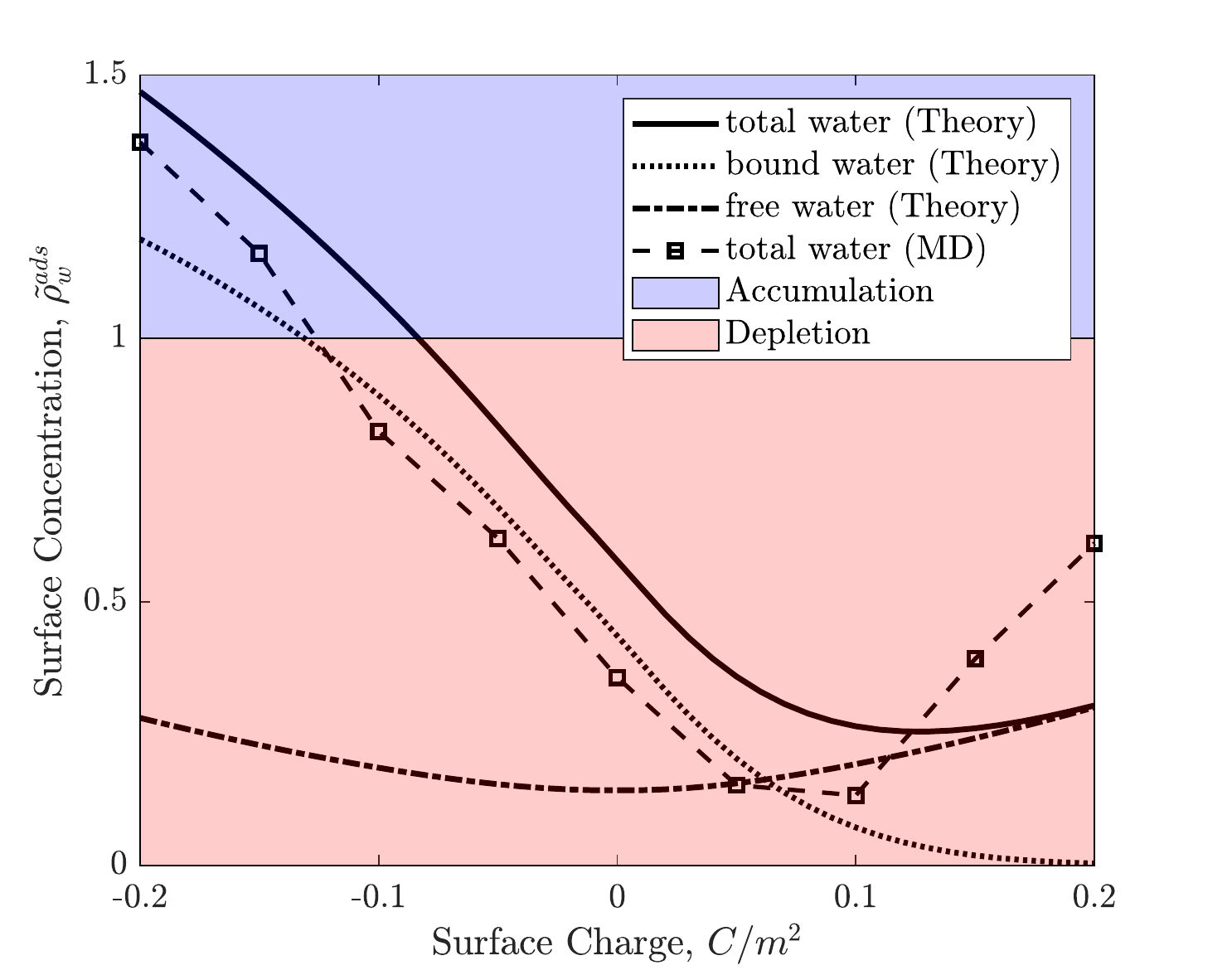}
\end{subfigure}
\begin{subfigure}[B]{0.49\textwidth}
    \includegraphics[clip, trim=3.5cm 8cm 3cm 8cm, width=1\textwidth]{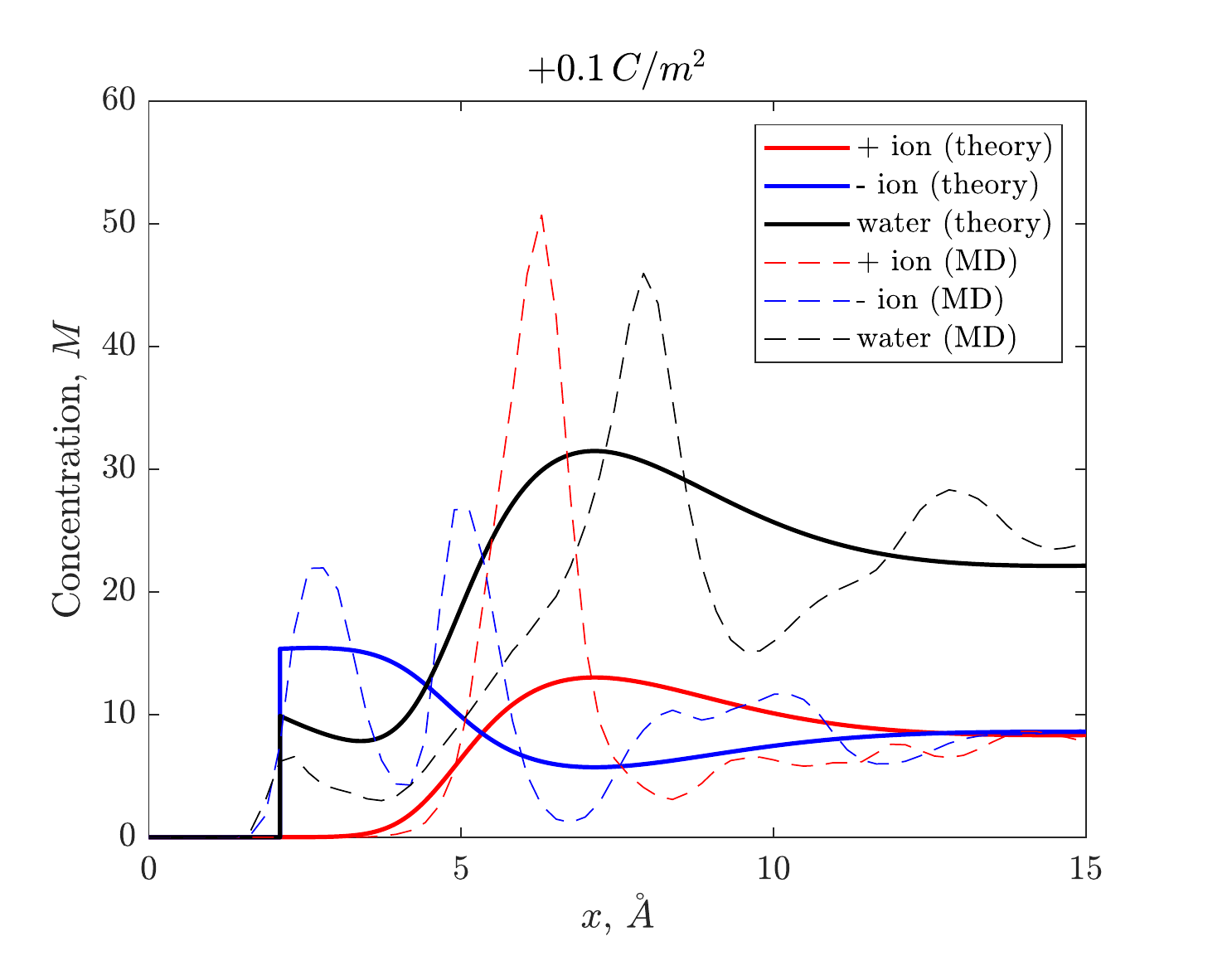}
\end{subfigure}
\begin{subfigure}[B]{0.49\textwidth}
    \includegraphics[clip, trim=3.5cm 8cm 3cm 8cm, width=1\textwidth]{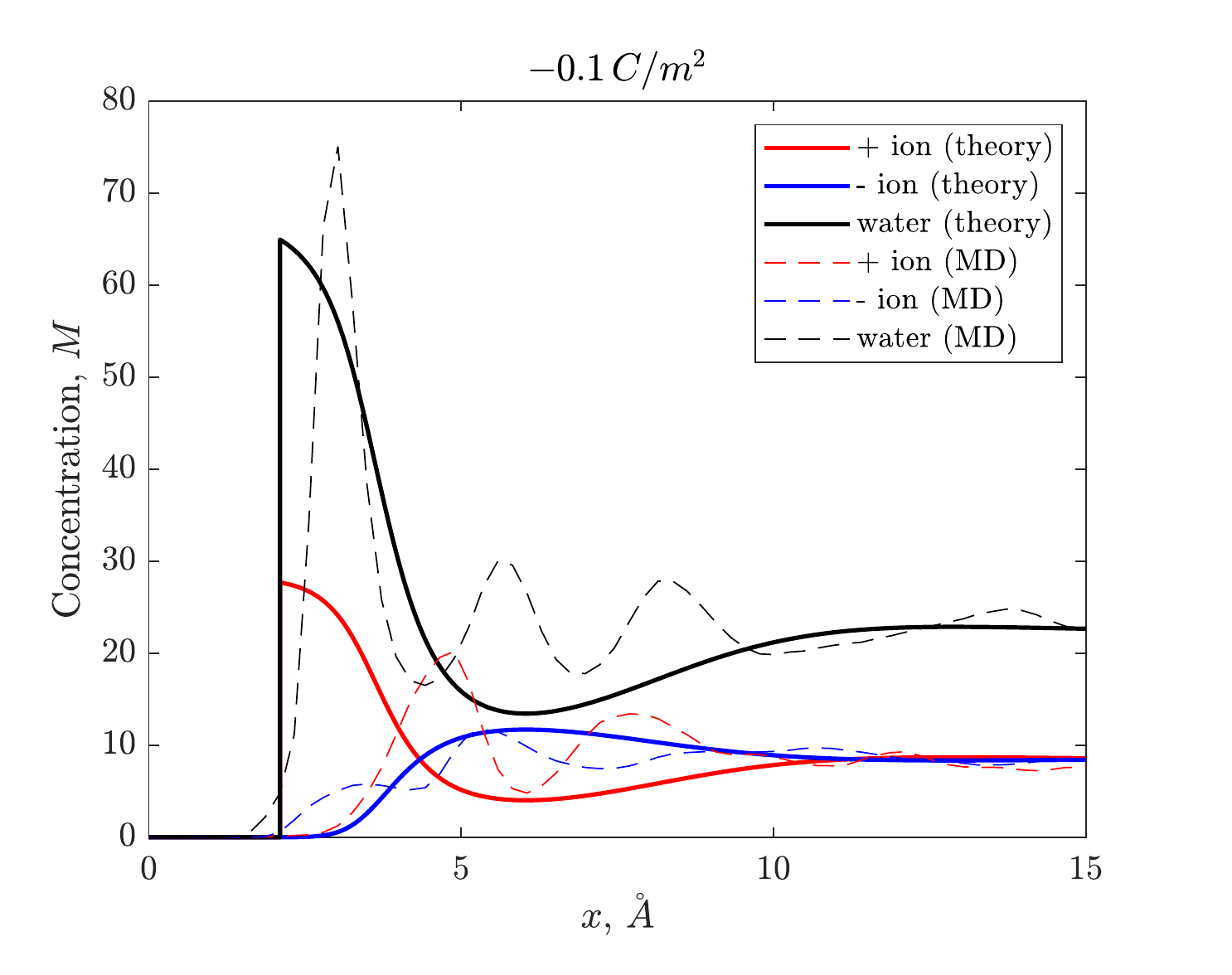}
\end{subfigure}

\caption{(A) Surface excess concentration isotherms for all species. (B) The interfacial concentration of water integrated within 5 $\AA$ of the electrode surface. The concentration profiles of species in 21m LiTFSI next surfaces with charge (C) $+0.1 C/m^2$ and (D) $-0.1 C/m^2$}
\label{fig:esorb_liotf}
\end{figure*}

Here we take a similar approach. One complication is that in our system, there is a large asymmetry in the size of cations and anions. MD simulations (see fig. \ref{fig:esorb_liotf}, for example) show asymmetric overscreening in when when the surface is charged positively or negatively. We posit that this observation implies that $l_c$ is a function of the size of the counter-ion attracted to the charged surface. For LiTFSI (and LiOTF), we assumed that $l_c$ is proportional to the counter-ion radii, $a_\pm$, with proportionality constant, $\alpha$. Thus, for $\sigma>0$, $l_c = \alpha a_-$ and for $\sigma<0$, $l_c = \alpha a_+$ 
\par
Thus, we have fully parameterized our model from MD simulations. Although we have outlined the parameterization for LiTFSI, we performed this procedure for LiOTF, as well. The determined parameters are written in table \ref{tbl:params}.

\begin{table}[!hbt]
\centering
\label{tbl:hpc_3d}
\begin{tabular}{|c|c|c|}
\hline
Parameter & LiTFSI & LiOTF\\
\hline
$\Phi_B$                 & 0.81   & 0.83 \\
$n_w^{\text{bound}}$     & 2.5    & 2.0  \\
$\xi_+$                  & 4.68   & 4.37  \\
$\xi_-$                  & 14.85  & 8.02  \\
$p_w/p_w^0$              & 2.7    & 3.0  \\
$\varepsilon_s$          & 10.1   & 8.2  \\
$\alpha$                 & 2.0    & 4.0   \\
$a_-$/$\AA$              & 3.53   & 2.87 \\
$a_+$/$\AA$              & 2.41   & 2.32 \\
\hline
\end{tabular}
\caption{Summary of Model Parameters}
\label{tbl:params}
\end{table}
\newpage
\subsection{Model Results for LiOTF}
\newpage
\bibliography{WiSE_Paper.bib}
\end{document}